\documentclass{ifacconf}

\usepackage{natbib}

\usepackage[T1]{fontenc}
\usepackage[utf8]{inputenc}

\usepackage{graphicx,amsmath,url,amssymb,mathtools,steinmetz,bm}
\usepackage[nameinlink,capitalise]{cleveref}
\crefname{appsec}{Appendix}{Appendices}
\crefname{equation}{}{}
\usepackage{autonum} 
\newcommand{\retainlabel}[1]{\label{#1}\sbox0{\ref{#1}}}
\usepackage{bbold}
\usepackage{subcaption}

\usepackage{siunitx}
\DeclareSIUnit\perunit{p.u.}
\DeclareSIUnit\voltampere{VA}

\usepackage[usenames, dvipsnames, monochrome]{color}

\newcommand{\steady}{\ensuremath{{*} }}
\newcommand{\T}{\textit{\textsf{T}}}
\newcommand{\inv}{\ensuremath{{-1}}}
\newcommand{\real}{\ensuremath{\mathbb{R}}}

\newcommand{\complex}{\ensuremath{\mathbb{C}}}
\newcommand{\neighbour}{\ensuremath{\mathcal{N}}}

\DeclareMathOperator{\Real}{\ensuremath{\mathrm{Re}}}
\DeclareMathOperator{\Imag}{\ensuremath{\mathrm{Im}}}
\DeclareMathOperator{\diagf}{\ensuremath{\mathrm{diag}}}
\DeclareMathOperator{\sign}{\ensuremath{\mathrm{sgn}}}

\newcommand{\Ybus}{\ensuremath{\bm{Y}}}
\newcommand{\modelM}{\ensuremath{\bm M}}
\newcommand{\modelT}{\ensuremath{\bm T}}
\newcommand{\modelE}{\ensuremath{\bm E}}
\newcommand{\subq}{\ensuremath{{\smash{q}}}}
\newcommand{\shunt}{\ensuremath{\mathrm{sh}}}
\newcommand{\bslash}{\ensuremath{\hat{b}}}

\usepackage{tikz}
\newcommand*\circled[1]{%
	\tikz[baseline=(C.base)]\node[draw,circle,inner sep=1.2pt,line width=0.2mm,](C) {\scriptsize #1};}
\newcommand*\Myitem{%
	\stepcounter{enumi}\item[\circled{\theenumi}]}

\newtheorem{rem_}{Remark}
\Crefname{rem_}{Remark}{Remarks}

\Crefname{defi}{Definition}{Definitions}
\newtheorem{asm}{Assumption}
\Crefname{asm}{Assumption}{Assumptions}
\usepackage{microtype}
\usepackage{placeins}
\usepackage{tikz}
\usepackage{tikzpagenodes}

\begin{document}
\begin{frontmatter}

\title{
	Control Limitations due to Zero Dynamics in a Single-Machine Infinite Bus Network%
	\thanksref{footnoteinfo}} 

\thanks[footnoteinfo]{This work was done under the PhD program in the digitalization of electric
	power engineering, School of Electrical Engineering and Computer Science,
	KTH Royal Institute of Technology, Sweden.
	\\
	The work was supported in part by the Knut and Alice Wallenberg Foundation,
	the Swedish Research Council, and the Swedish Foundation for Strategic
	Research.
\\
$
\smash{
\mathrlap{
\begin{matrix}
\\
\\
\text{\normalsize\tt This work has been submitted to IFAC for possible publication.}
\end{matrix}}
}
$
}

\author[First]{Joakim Bj\"ork} 
\author[First]{Karl Henrik Johansson} 

\address[First]{
	School of 
	Electrical Engineering and Computer Science,\\ KTH Royal Institute of Technology, 
   Stockholm, Sweden \\(e-mail: joakbj@kth.se, kallej@kth.se).}

\begin{abstract}                
In this work, fundamental control limitations for rotor angle stability are considered. Limitations are identified by characterizing open-loop transfer function zeros for input-output combinations of certain power system configurations. Of particular interest are non-minimum phase (NMP) zeros that limit the achievable performance of the closed-loop system. By studying a single-machine infinite bus power system model, analytic conditions for the presence of NMP zeros are derived. They are shown to be closely linked to the destabilizing effect of automatic voltage regulators (AVRs). Depending on the control loop, it is found that NMP zeros may persist in the system even if the closed-loop system is stabilized through feedback control. A simulation study shows that NMP zeros introduced by AVR limit the achievable performance and stabilization using feedback control.
\end{abstract}


\end{frontmatter}


\section{Introduction}
It is well known that stressed power systems experience unstable electromechanical modes. 
The instability can often be attributed to the automatic voltage regulators (AVRs) which are needed to maintain synchrony following large disturbances. 
Eigenvalue analysis is used to identify poles associated with the poorly damped electromechanical modes and to design power oscillation damping (POD) controllers, such as power system stabilizers (PSSs), to move these poles into the left half-plane~\citep{KundurPowerSystemStability1994}.

The study of fundamental limitations in filtering and control design reaches back to the ground breaking work of Bode in the 1940's, as subsequently published in  \citep{bodeNetworkAnalysisFeedback1945}. In this paper, we consider the feedback control limitations associated with non-minimum phase (NMP) zeros. 
With increasing feedback gain, the closed-loop poles tends to the position of the open-loop zeros. NMP zeros therefore introduce a limitation on the achievable performance of the closed-loop system.
 For an overview of control limitations associated with NMP zeros see for instance \cite{seronFundamentalLimitationsFiltering1997}. 
When designing PSS we typically only care about a bandwidth window around poorly damped poles. Thus only NMP zeros close to the considered poles impose limitations on the closed-loop system.
Since zero positions highly depend on the operating condition, they need to be carefully analyzed. An example is the modulation control of the Pacific DC Intertie in the 1970's. The modulation control considerably improved stability of the North-South inter-area mode in the western North American power system. However, using local ac power flow as feedback signal, the open-loop system showed a NMP zero that caused the modulation to introduce a \SI{0.7}{\hertz} oscillation under certain operating conditions~\citep{cresapOperatingExperienceModulation1978}. This was one of the primary reasons that the control method eventually got rejected~\citep{trudnowskiPDCIDampingControl2013}.

Rigorous numerical case studies are required to identify troublesome zero dynamics and to gain insight into the control problem at hand~\citep{jonesZeroDynamicsRobust1999}. 
Today, efficient methods are available to identify transfer function zeros even in large power system models~\citep{martinsEfficientMethodsFinding1992}.  
In \citep{dominguez-garciaControlSignalSelection2013}, trade-offs between local measurements and wide-area measurement are studied in the general control configuration. Limitations due to NMP zeros and time-delays are taken into account to identify suitable input-output pairs.
Resorting only to a numerical solution loses valuable insight into the problem at hand. In~\citep{smedUtilizingHVDCDamp1993} it is shown analytically how the location of controlled power injections in the system affects the potential of POD control. 
In~\citep{zhangInterconnectionTwoVery2011} a simplified analytical model is used to study the connection of a  voltage-source converter (VSC) based high-voltage dc  link to a weak ac system. Open-loop NMP zeros are found to approach the origin with increasing load angles. This causes bandwidth limitations on the voltage control of the VSC, suggesting that high dc capacitance is needed when connecting to weak grids. \cite{bjorkInfluenceSensorFeedbackunpublished} studies POD by controlling active power injections between two oscillating regions. It is shown that a reduction of transient stability is unavoidable if POD control is designed using local frequency measurements.

The contribution of this paper is to identify fundamental control limitations for POD. This is done by characterizing the open-loop transfer function zero dynamics  for different input-output combinations in the system. To obtain useful analytical results, the focus is on a single-machine infinite bus (SMIB) model. For this model, it is explicitly shown how AVR destabilizes the electromechanical mode. For some transfer functions, NMP zeros are found to be caused by interaction with the AVR.
Depending on the control-loop, it is shown that the NMP zeros persist in the system even if the closed-loop system is stabilized.
The analysis gives insight into where in the system NMP zeros are likely to occur, and where they may impose trouble for POD control design.

The remainder of this paper is organized as follows.
In \Cref{sec:model} a linearized SMIB model is presented. In \Cref{sec:ctrllimSMIB} control limitations are derived and in \Cref{sec:simulation} the results are verified on a detailed nonlinear power system model. \Cref{sec:Conclusions} concludes the work.

\section{Model}
\label{sec:model}
We consider a nonlinear differential algebraic power network model on the form
\begin{subequations}
	\begin{align}
	\dot x(t) &= f\left(x(t),\gamma(t),u(t)\right) \label{eq:diffeq} \retainlabel{eq:diffeq}
	\\
	0 &= g\left(x(t),\gamma(t),u(t)\right)  \label{eq:algeq}
	\\
	y(t) &= h\left(x(t),\gamma(t),u(t)\right)  \label{eq:outputeq} \retainlabel{eq:outputeq}
	\end{align}
\end{subequations}
where $x(t)\in\real^{n_x}$ constitute the states, $\gamma(t)\in\real^{n_\gamma}$ are time-varying parameters, $u(t)\in\real^{n_u}$ external inputs, and $y(t) \in \real^{n_y}$ some outputs of the system. 

For the analysis, the system is linearized at a stationary operating point $x(t) =x^\steady$, $\gamma(t) =\gamma^\steady$, and $u(t) =u^\steady$, resulting in the linear state-space model
\begin{equation}
\label{eq:state-space-general}
\begin{aligned}
\dot x(t) &= Ax(t) + Bu(t)\\
y(t) &= Cx(t) + Du(t).
\end{aligned}
\end{equation}
Since the time constants in \eqref{eq:state-space-general} depend on the current operating point, the model can only be considered accurate for small deviation from this point.
\subsection{Network Model}
\label{section_netwrok_model}
Consider a power network and let the system voltages be given by
\begin{equation}
\label{eq:voltage1}
Ue^{j\varphi}  =  \begin{bmatrix}
U_1e^{j\varphi_1},U_2e^{j\varphi_2},\ldots,  U_Ne^{j\varphi_N}
\end{bmatrix}^{\smash{\T}} \in \complex^N
\end{equation}
where $U\in\real^N$ and $\varphi\in\real^N$ are voltage amplitudes at system nodes and the phase angles relative to a constant reference frame rotating at nominal frequency $f_\text{nom}$. Typically $f_\text{nom} = 50$ or \SI{60}{\hertz}.
The impedance between two nodes is given by
\begin{equation}
z_{ik} = r_{ik} + jx_{ik}
\end{equation}
and corresponding admittance is
\begin{equation}
y_{ik} = 1/z_{ik} =  g_{ik} - jb_{ik}.
\end{equation}
Nodes are interconnected over a network described by the admittance matrix $Y\in\complex^{N\times N}$
with elements
\begin{equation}
\label{eq:netwrokAdmit_0}
Y_{ik} = -y_{ik},  \quad \text{and} \quad Y_{ii} = y_{ii} =  y_i +\sum_{\mathclap{k\in\neighbour_i}} y_{ik}
\end{equation}
where $y_i = r_i-jb_i$ is the shunt admittance at node $i$ and $\neighbour_i$ is the set of nodes directly connected to node $i$.
Power injected at the nodes are given by 
\begin{equation}
\label{eq:complexPowerInjection}
S = P + j Q = \diagf (Ue^{j\varphi}) \bar Y Ue^{-j\varphi}\in \complex^{N},
\end{equation}
where $\bar Y$ is the complex conjugate of $Y$.
Active and reactive power injected at node $i$ give the two algebraic equations
\begingroup\makeatletter\def\f@size{9}\check@mathfonts
\def\maketag@@@#1{\hbox{\m@th\normalsize\normalfont#1}}%
\begin{align}
P_i &= g_{ii} U_i^2 + \sum_{\mathclap{k\in\neighbour_i}} U_iU_k\left[b_{ik} \sin(\varphi_i-\varphi_k)-g_{ik}  \cos(\varphi_i-\varphi_k)\right]\notag \\
Q_i &= {b_{ii}} U_i^2 - \sum_{\mathclap{k\in\neighbour_i}} U_iU_k\left[b_{ik} \cos(\varphi_i-\varphi_k)+g_{ik} \sin(\varphi_i-\varphi_k)\right].
\end{align}
\endgroup
For the analysis it is convenient to write \eqref{eq:complexPowerInjection} as $S = \Ybus \mathbb{1}$, where $\mathbb{1}\in\real^N$ is a vector of ones and $\Ybus\in\complex^{N\times N}$ is a \emph{weighted} admittance matrix with elements
\begin{equation}
\label{def:weightAdmit}
\Ybus_{ik} = -\bar y_{ik}U_iU_ke^{j(\varphi_i-\varphi_k)},  \quad \text{and} \quad \Ybus_{ii} = \bar y_{ii} U_i^2.
\end{equation}	

We partition \eqref{eq:voltage1} as
\begin{equation}
\label{eq:voltages_split}
Ue^{j\varphi}  =\big[
\underbrace{(Ee^{j\delta})^\T}_
{\smash{\mathclap{\text{Dynamic nodes.}}\hspace{7mm}}}, 
\underbrace{(Ve^{j\theta})^\T}_
{\smash{\mathclap{\hspace{18mm}\text{Algebraic network nodes.}}}}
\big]^\T \in\complex^N,
\end{equation}
to differentiate between nodes where voltages $Ee^{j\delta}\in\complex^{n_\delta}$ are determined by differential equations \eqref{eq:diffeq}, and nodes where voltages $Ve^{j\theta}\in\complex^{n_\theta}$ are determined by algebraic equations \eqref{eq:algeq}.
\subsection{One-Axis Synchronous Machine Model}
For study of electromechanical dynamics, the synchronous machines in the system are often described using the one-axis model, with notation from \cite{sauerPowerSystemDynamics1998},
\begin{equation}
\begin{aligned}
\dot \delta &= \omega\\
\label{eq:one-axis1}
M \dot \omega &= -\frac{1}{x_{d}'} E_{q}'V \sin(\delta-\theta)+ D_m\omega + P_{m} 
\\
T_{do}' \dot{E}_{q}' &= -\frac{x_{d}}{x_{d}'}E_{q}' + \frac{x_{d}-x_{d}'}{x_{d}'}V\cos(\delta-\theta)+ E_{f},
\end{aligned}
\end{equation}
where state variables $\delta$, $\omega$, and $E_q'$ represent the rotor phase angle, rotor speed deviation from nominal speed, and $q$-axis transient voltage, respectively.
Parameter $M$ represent the machine inertia, $D_m$ a small non-negative machine damping constant, $T'_{do}$ the $d$-axis transient open-circuit time constant, and   $x_{d}'$ and $x_{d}$ the $d$-axis transient reactance and synchronous reactance respectively.\footnote{The prime notation is used to emphasize that the model assumes parameters linearized at a fixed speed and that the model is accurate only for a transient time period. Faster \emph{subtransient} dynamics are often noted with double prime. In the one-axis model, the subtransient and $d$-axis dynamics have been neglected by setting the corresponding time constants $\smash{T_{do}''=T_{qo}''=T_{qo}'=0}$ \citep{sauerPowerSystemDynamics1998}.
}~External inputs $P_m$ and $E_f$ are the mechanical power from the turbine and the field voltage. Variables $V$ and $\theta$ represent the voltage amplitude and phase angle at the adjacent network node. The node adjacent to the machine node will typically be referred to as the machine terminal. The total series reactance between the machine node and the machine terminal include transformers, line reactances etc.
\begin{figure}[t]
	\centering
	\includegraphics[scale=0.9]{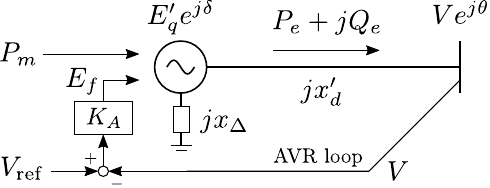} 
	\caption{One-axis synchronous machine with AVR.}
	\label{fig:oneMachine}
\end{figure}

Let $P_{e}$ and $Q_{e}$ be the active and reactive power exported from the machine node to the network as shown in \cref{fig:oneMachine}.
The second line in \eqref{eq:one-axis1} corresponds to {$M \dot \omega_ = -P_{e} + D_m\omega + P_{m}$}, i.e., rotor angular acceleration depends on the active power balance at the machine node. Similarly the third line is a function of the reactive power. 
Let $x_{\Delta } = x_{d}-x_{d}'$, $b_d' = 1/x_d'$, and $b_\Delta = 1/x_\Delta$.
The third line in \eqref{eq:one-axis1} can then be rewritten as
\begin{equation}
\label{eq:one-axis_thirdline2}
\label{def:voltage_dyn}
{T_{do}'}{b_{\Delta }} \dot{E}_{q}' = 
-{b_{\Delta }}E_{q}' \! 
\underbrace
{-{b_{d}'}E_{q}' + {b_{d}'}V\cos(\delta-\theta)}
{\smash{-Q_{e}/E_{q}'}} 
+ {b_{\Delta }}E_{f}.
\end{equation}	
Note that $b_{\Delta }$ represents the shunt susceptance at the machine node. Thus, we can describe the machine node as a dynamical node connected to an algebraic network node. 

\subsection{Excitation Control of Synchronous Machine}
The excitation system  performs control and protective functions essential to satisfactory performance of the power system by controlling the field voltage, $E_f$. 
High-speed excitation systems with AVRs are commonly installed at generators as it is by far the most effective and economical method to improve transient stability \citep{KundurPowerSystemStability1994}. 
AVRs are typically modeled using the first-order model 
\begin{equation}
\label{eq:AVR_original}
T_e \dot E_f = -E_f  +K_A (V_\text{ref}-V),
\end{equation}
where $V$ is typically measured at the machine terminal. For the purpose of our analysis, however, it is assumed that the fast dynamics of the excitation system can be neglected, so
\begin{equation}
\label{eq:AVR}
E_f =  K_A (V_\text{ref}-V),
\end{equation}
as shown in \cref{fig:oneMachine}.

\subsection{Linearized Multi-Machine Model}
\label{sec_inputoutputmatrices}
\begin{figure}[t]
	\centering
	\includegraphics[width = \linewidth]{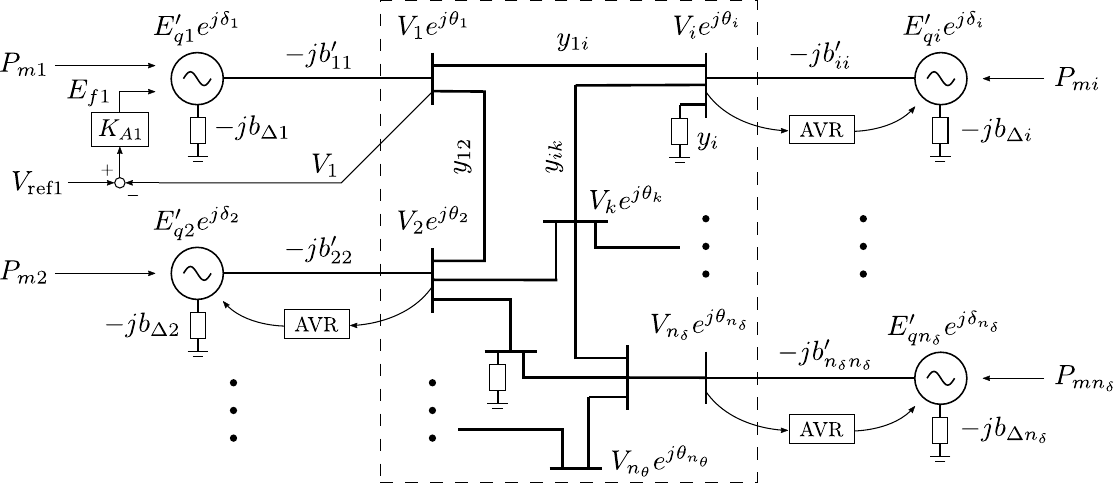} 
	\caption{Multi-machine power system with $n_\delta$ machines interconnected over a network with $n_\theta\geq n_\delta$ nodes.}
	\label{fig:multiMachine}
\end{figure}

Consider a multi-machine power system, as shown in \cref{fig:multiMachine}, with $n_\delta$ machines represented using the one-axis model \eqref{eq:one-axis1}. Let state variables $x=[\delta^\T,\omega^\T,E_{\subq}'^\T]^\T \in \real^{3n_\delta}$ represent the generator states and algebraic variables $\gamma = [\theta^\T,V^\T]^\T \in \real^{2 n_\theta}$ voltages at the $n_\theta$ network nodes. 

Partition the weighted network admittance  matrix \eqref{def:weightAdmit} as
\begin{equation}
\label{eq:Ybus_appendix}
\Ybus = \left[
\begin{array}{c|c}
\Ybus_{\delta\delta} & \Ybus_{\delta\theta} \\
\hline
\Ybus_{\theta\delta} & \Ybus_{\theta\theta}
\end{array}\right] \in \complex^{N\times N}
\end{equation}
where $\Ybus_{\delta\delta}\in\complex^{n_\delta\times n_\delta}$. If the machine terminals are modeled, as shown in \cref{fig:multiMachine}, the corresponding \emph{unweighted} admittance matrix \cref{eq:netwrokAdmit_0} is given by $Y_{\delta\delta} = -j\diagf(b_d'+b_\Delta) $. 
Network matrix $\Ybus_{\theta\theta}\in\complex^{n_\theta \times n_\theta}$ connects network nodes where voltages are determined by the algebraic equation \eqref{eq:algeq}. Off-diagonal blocks $\Ybus_{\delta\theta}\in\complex^{n_\delta \times n_\theta}$ and $\Ybus_{\theta\delta} \in\complex^{n_\theta \times n_\delta}$ models the connection between the machine and network nodes. 
Constant power inputs are assumed to be zero at the algebraic network  nodes, whereas
constant impedance loads can be incorporated as shunt elements in \eqref{eq:netwrokAdmit_0}.

Following the modeling above and excluding the AVR, the state matrix of \eqref{eq:state-space-general} becomes $A_0=$ 
\begingroup\makeatletter\def\f@size{9.5}\check@mathfonts
\def\maketag@@@#1{\hbox{\m@th\normalsize\normalfont#1}}%
\begin{equation}
\begin{bmatrix}
0 &I& 0
\\
-\bm M^\inv \Imag(\Ybus_A) &-\bm M^\inv \bm D_m &
-\bm M^\inv \Real(\Ybus_A+2\Ybus_\text\shunt)\modelE^\inv
\\
\bm \modelT^\inv \Real(\Ybus_A) & 0&
-\bm \modelT^\inv \Imag(\Ybus_A+\Ybus_\text\shunt) \modelE^\inv
\end{bmatrix}\!,
\label{app:statmatrix}
\end{equation}
\endgroup
where $0$ and $I$ are appropriately sized zero and identity matrices, $\bm M = \diagf(M)$, $\bm D_m = \diagf(D_m)$, $\bm \modelT = \diagf(T_{do}'b_\Delta E_{\subq}'^*)$, and $\modelE = \diagf( E_{\subq}'^* ) \in\real^{n_\delta \times n_\delta}$. The network matrix $\Ybus_A$ is obtained from  $\Ybus_\text{red} = \Ybus_{\delta\delta} -\Ybus_{\delta\theta} \Ybus_{\theta\theta}^\inv\Ybus_{\theta\delta}$ and  $\Ybus_\text\shunt = \diagf(\Ybus_\text{red}\mathbb{1})$, as $\Ybus_A = \Ybus_\text{red} - \Ybus_\text\shunt$,
evaluated around a steady-state operating point. Note that $\Ybus_\shunt$ contains the power injected by turbines and excitation system at the machine nodes.

Considering constant power inputs $u=[P^\T,Q^\T]^\T \in \real^{2 n_\theta}$  and outputs $y  = \gamma$ at network nodes, we have
input matrix, excluding AVR and with $u^*=0$,
\begin{align}
\label{app:inputmatrix}
B_0 &= \begin{bmatrix}
0 & 0
\\
\modelM^\inv \Real(\Ybus_B) & - \modelM^\inv \Imag(\Ybus_B)
\\
\bm \modelT^\inv \Imag(\Ybus_B) &
\bm \modelT^\inv \Real(\Ybus_B) 
\end{bmatrix},
\intertext{
	output matrix 
}
\label{app:outputmatrix}
C &=\begin{bmatrix}
\Real(\Ybus_C) & 0 & -\Imag(\Ybus_C) \modelE^\inv
\\
\bm V \Imag(\Ybus_C) & 0 & \bm V \Real(\Ybus_C) \modelE^\inv
\end{bmatrix},
\intertext{
	and
	direct feed-through matrix
}
\label{app:feedthroughmatrix}
D &= \begin{bmatrix}
\Imag(\Ybus_D) & \Real(\Ybus_D)
\\
-\bm V \Real(\Ybus_D) & \bm V \Imag(\Ybus_D)
\end{bmatrix},
\end{align}
where $\Ybus_D = -\Ybus_{\theta\theta}^\inv$, $\Ybus_B = \Ybus_{\delta\theta} \Ybus_{D}$, $\Ybus_C =\Ybus_{D} \Ybus_{\theta\delta}$, and $\bm V = \diagf(V^*) \in\real^{n_\theta \times n_\theta}$. 

With AVR modeled using \eqref{eq:AVR}, the state matrix of \eqref{eq:state-space-general} is instead given by 
\begin{equation}
\label{eq:app_A_AVR_addition}
\begin{aligned}
A = A_0 - \bm K_A \bm T_{do}'^\inv
\end{aligned}
\begin{aligned}
\begin{bmatrix}
0 & 0 &0
\\
0 & 0 &0 
\\
\bm V \Imag(\Ybus_C) & 0 & \bm V \Real(\Ybus_C) \modelE^\inv 
\end{bmatrix},
\end{aligned}
\end{equation}
and the input matrix by
\begin{equation}
\label{eq:app_B_AVR_addition}
B = B_0 - \bm K_A \bm T_{do}'^\inv 
\begin{bmatrix}
0 & 0
\\
0 & 0
\\
-\bm V \Real(\Ybus_D) &
\bm V \Imag(\Ybus_D) 
\end{bmatrix},
\end{equation}
where $\bm K_A \bm T_{do}'^\inv = \diagf(K_A) \diagf(
T_{do}')^\inv$.

\subsection{Single-Machine Infinite Bus Model}
\begin{figure}[t]
	\centering
	\includegraphics[scale=0.9]{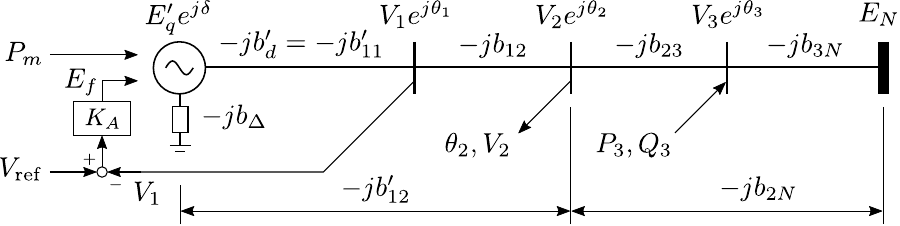} 
	\caption{SMIB model with three network nodes. }
	\label{fig:SMIB}
\end{figure}

The single-machine infinite bus (SMIB) model shown in \cref{fig:SMIB} is commonly used for analyzing generators connected to the grid. The machine is connected to a strong (infinite) bus that represent the rest of the system. Here, the voltage $E_N$ can be considered as a constant voltage rotating at the nominal system frequency. 
We introduce three network nodes as shown in \cref{fig:SMIB}.
The system voltages are given by
\begin{equation}
Ue^{j\varphi} = \big[
{E_{\subq}'e^{j\delta} \ E_{N}}
\ 
{V_1e^{j\theta_1} \ V_2e^{j\theta_2} \ V_3e^{j\theta_3}}
\big]^\T,
\end{equation}
where $V_1$ is measured for the AVR. 
The linearized state-space model \eqref{eq:state-space-general} is obtained as above with the unforced dynamics for the states $x = [\delta,\omega,E_{\subq}']^\T$ given by\footnote{Dynamics at the infinite bus are truncated since $\dot E_N=\dot \delta_N = 0$.}
\begin{equation}
\label{eq:Amat}
\dot x-\dot x^* = A(x-x^*) =  \begin{bmatrix}
0& 1& 0 \\
-a_{21} & -a_{22} & -a_{23} \\
-a_{31} & 0 & -a_{33}
\end{bmatrix}(x-x^*)
\end{equation}
with elements
\begin{align}
a_{21} &= \frac{b_\Sigma}{M} E_{\subq}'^\steady E_N \cos\delta^*, \ \ 
a_{22} = \frac{D_m}{M} , \ \
a_{23} = \frac{b_\Sigma}{M} E_N \sin\delta^*, \notag
\\
a_{31} &= \frac{b_\Sigma}{T_{do}' b_\Delta} E_N \sin\delta^* - \frac{K_A}{T_{do}'}\beta_1 E_{\subq}'^\steady\sin\varepsilon^*_1, \notag
\\
a_{33} &= \frac{b_\Delta+b_\Sigma}{T_{do}' b_\Delta} +\frac{K_A}{T_{do}'}\beta_1 \cos\varepsilon^*_1, \rule{4cm}{0pt}  \raisetag{35pt}\label{eq:Avariables}
\end{align}
where $b_\Sigma$ is the total series susceptance between the machine node and the infinite bus.
Note that element $a_{31}$ and $a_{33}$ are affected by the AVR using measurements at the machine terminal. 
At network nodes $i=1,2,3$,
\begin{equation}
\label{eq:SMIBout1}
\begin{bmatrix}
\theta_i-\theta_i^* \\
V_i-V_i^* 
\end{bmatrix}
= \beta_i\begin{bmatrix}
\frac{E_q'^*}{V_i^*} \cos \varepsilon_i^* & 0 & \frac{1}{V_i^*}\sin \varepsilon_i^*
\\
- E_\subq'^* \sin \varepsilon_i^* & 0 & \cos \varepsilon_i^*
\end{bmatrix} (
x-x^*
),
\end{equation}
where $\beta_i = b_{1i}'/(b_{1i}'+b_{iN}) \in[0,1]$ is the relative electric position of the network node and $\varepsilon_{i}^* = \delta^*-\theta_i^*$. 
The input matrix \cref{app:inputmatrix} can be derived similar to \eqref{eq:SMIBout1}.
With AVR, a direct feed-through between the input and voltage measurement at the machine terminal are introduced as shown in \eqref{eq:app_B_AVR_addition}. The direct feed-through \eqref{app:feedthroughmatrix} between nodes are given in \Cref{app:direct_feedthorugh}.

\section{SMIB Control Limitations}
\label{sec:ctrllimSMIB}
Consider the SMIB model in \cref{fig:SMIB}. At the machine terminal, $V_1$ is measured for the AVR. We will first show how interaction between the electromechanical and voltage dynamics have a destabilizing effect on the electromechanical mode. Then we study the control performance limitations in stabilizing this mode.
This is done by studying the open-loop zeros in the SISO transfer function $G_{yu}(s)$ from inputs $u=P_3$~or~$Q_3$ at a control bus to  measurements $y=\theta_2$~or~$V_2$ at a measurement bus.\footnote{For PSS, the control bus coincide with the machine node.}~To limit the number of possible scenarios, the following standing assumption is made.
\begin{asm}
	\label{asm:uniform_power_flow}
	Active power flows in a uniform direction between the machine node and the infinite bus. Load angles $\varepsilon^*_i=\delta^*-\theta^*_i$, $i = 1,2,3$, therefore have the same sign. Network nodes may coincide or be placed in any order between the machine node and the infinite bus.
\end{asm}
\subsection{AVR Influence on Stability}
\label{sec:AVRstab}
In this section we consider the influence of the AVR on the dynamics. 
A common simplifying assumption is that the load angle $\delta^*\approx \theta_i^* \approx 0$.  
Under this assumption $a_{23} = a_{31} = 0$ and \eqref{eq:Amat} has three eigenvalues: $\lambda_{1,2} \approx -a_{22}/2 \pm j \sqrt{a_{21}}$ (assuming $a_{21} \gg a_{22}$\footnote{With $\omega$ in \si{\radian\per\second}, typically $D_m \in [0,\,3/(2\pi f_\text{nom})]$. Thus $a_{21} \gg a_{22}$.})
and $\lambda_3 = -a_{33}$ associated with $[\delta,\omega]$ and $E_{\subq}'$ respectively. 

Now as $\delta \neq 0$ the voltage mode $\lambda_{3}$ will start to interact with the electromechanical mode $\lambda_{1}$, $\lambda_{2}$. If $K_A = 0$, or if $\theta_1^* = \delta^*$, then since $\sign(a_{31}) = \sign(a_{23})$ the eigenvalues will attract each other, thus stabilizing the electromechanical mode. However, usually with AVR, $K_A \gg 0$ and $\theta_1^*\neq \delta^*$. In this case $\lambda_3$ moves further into the left half-plane and is mostly unaffected by the interaction. This does not hold true for the electromechanical mode, however. We notice that if
\begin{equation}
\label{eq:destabCond}
\left|\frac{b_\Sigma}{b_\Delta} E_N \sin\delta^*\right| >
\left|K_A\beta_1 E_{\subq}'^\steady\sin\varepsilon_1^*\right|,
\end{equation}
is violated, then $a_{31}$ changes sign. Therefore, the eigenvalues will repel each other. The interaction thus destabilizes the electromechanical mode. 

The same conclusion can be draw from studying the characteristic polynomial of \eqref{eq:Amat}:
\begingroup\makeatletter\def\f@size{10}\check@mathfonts
\def\maketag@@@#1{\hbox{\m@th\normalsize\normalfont#1}}%
\begin{multline}
\label{eq:characteristic_eq}
p(s) = s^3 + (a_{22}+a_{33})s^2 + (a_{22}a_{33}+a_{21})s \\+ a_{21}a_{33} - a_{23}a_{31}.
\end{multline}
\endgroup
This polynomial is stable if
\begin{multline}
\beta_1  K_A \left(\cos\delta^*\cos\varepsilon_1^*+
\sin\delta^*\sin\varepsilon_1^*\right) 
\\
+\frac{(b_\Delta+b_\Sigma)}{b_\Delta}\left(\cos\delta^* - \frac{E_N}{E_\subq'^*}\sin^2\delta^*\right)
>0,
\end{multline}
which
holds true for reasonable load angles and large $K_A$,
and if
\begin{equation}
\label{eq:stabcriteria_with_damp}
{a_{22}(a_{22}a_{33} + a_{21}+a_{33}^2)} +a_{23}a_{31}>0.
\end{equation}
For the remainder of the section, we make the simplifying assumption
that the direct machine damping $D_m=0$.
Hence, $a_{22} = 0$, so the stability criterion becomes
\begin{equation}
\label{eq:stabcriteria}
a_{23}a_{31}>0,
\end{equation}
which is equivalent to \eqref{eq:destabCond}.

\subsection{Auxiliary Excitation Control (PSS)}
\label{sec:aux_exc_ctrl}
\begin{figure}[t]
	\centering
	\includegraphics[scale=0.9]{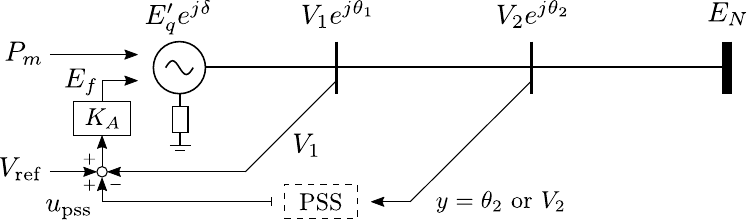} 
	\caption{SMIB model with PSS. Note that network node 2 may coincide directly with the machine node.}
	\label{fig:SMIBpss}
\end{figure}

A common way to improve stability is to combine AVR with PSS. 
With input $u_\text{pss}$ in \cref{fig:SMIBpss}, the transfer function to system states becomes
\begin{equation}
\label{eq:Ef_to_states}
(sI-A)^\inv \begin{bmatrix}
0\\ 0\\ K_A
\end{bmatrix} = \frac{K_A}{p(s)} \begin{bmatrix}
-a_{23}\\-sa_{23}\\s^2+a_{21}
\end{bmatrix},
\end{equation}
where $p(s)$ is given by \eqref{eq:characteristic_eq}. 
Note that zeros in \cref{eq:Ef_to_states} are unaffected by the AVR. 

Using \eqref{eq:SMIBout1},
the open-loop zeros of the transfer function from $u_\text{pss}$ to $\theta_2$, $G_{\theta_2,u_\text{pss}}(s)$ are
\begingroup\makeatletter\def\f@size{10}\check@mathfonts
\def\maketag@@@#1{\hbox{\m@th\normalsize\normalfont#1}}%
\begin{equation}
\label{eq:zerotheta}
\text{\normalsize$q_{1,2}^{(\theta_2,u_\text{pss})}$}
\! = \pm \sqrt{\frac{b_\Sigma}{M}E_{\subq}'^*E_N\left(\! \sin\delta^*\frac{\cos\varepsilon^*_2}{\sin\varepsilon^*_2}- \cos\delta^*\right)}.
\end{equation}
\endgroup
Since $\sign(\delta^*) = \sign(\varepsilon_2^*)$ and $|\delta^*| \geq |\varepsilon_2^*|$, the zeros are real if $\delta^*\neq 0$.  For PSS, rotor frequency are typically measured directly, in which case ${q_{\smash{\text{\tiny 1,2}}}^{(\theta_2,u_\text{pss})}}\rightarrow \pm \infty$ as $|\varepsilon_2^*| \rightarrow 0$. 

The open-loop zeros of $G_{V_2,u_\text{pss}}(s)$ are
\begingroup\makeatletter\def\f@size{10}\check@mathfonts
\def\maketag@@@#1{\hbox{\m@th\normalsize\normalfont#1}}%
\begin{equation}
\label{eq:zeroVolt}
\text{\normalsize$q_{1,2}^{(V_2,u_\text{pss})}$}
\! = \pm j \sqrt{\frac{b_\Sigma}{M}E_{\subq}'^*E_N\left( \! \cos\delta^*+ \sin\delta^* \frac{\sin\varepsilon^*_2}{\cos\varepsilon^*_2}\right)}.
\end{equation}
\endgroup
With direct machine damping $a_{22}>0$,
both zero pairs in \cref{eq:zerotheta,eq:zeroVolt} will move in the negative real direction with increased damping gain. 

\begin{rem_}[Observability]
	\label{rem:Vref_to_V2}
	The zeros \eqref{eq:zeroVolt} are very close to the undamped frequency of the electromechanical mode
	\begin{equation}
	\label{eq:OMEGA}
	\Omega = \sqrt{\frac{b_\Sigma}{M}E_{\subq}'^*E_N\cos\delta^*},
	\end{equation}
	making this an unsuitable loop to close.
\end{rem_}

\begin{rem_}[Power measurement]
	In the SMIB~model, active power measurement can be considered a special case of phase angle measurements, where $\varepsilon_2^* = \delta^*$ and $\beta_2 = b_\Sigma$ in \eqref{eq:SMIBout1}. By \eqref{eq:zerotheta}, this results in two zeros in the origin.
\end{rem_}

\subsection{Auxiliary Governor Control}
\label{sec:aux_gov_ctrl}
\cite{smedUtilizingHVDCDamp1993} show that active power injections close to the machine node are ideal for controlling the electromechanical mode. Assuming that the governor is fast enough so that its dynamics can be ignored,
the transfer function to system states becomes
\begin{equation}
\label{eq:Pm_to_states}
(sI-A)^\inv \begin{bmatrix}
0\\ 1\\ 0
\end{bmatrix} = \frac{1}{p(s)} \begin{bmatrix}
s+a_{33}\\s(s+a_{33})\\-a_{31}
\end{bmatrix}.
\end{equation}
The transfer function $G_{\theta_2,P_m}(s)$ has one open-loop zero
\begin{multline}
\label{eq:zeroPmtheta}
\text{\normalsize$q^{(\theta_2,P_m)}$}
\! = 
\frac{-1}{T_{do}' b_\Delta}\left(b_\Delta +
b_\Sigma \left[1-\frac{E_N}{E_q'^*} \frac{\sin \varepsilon_2^*}{\cos \varepsilon_2^*}
\sin\delta^*\right]
\right)
\\
-\frac{K_A}{T_{do}'}\beta_1 \left(\cos\varepsilon^*_1 + \frac{\sin \varepsilon_2^*}{\cos \varepsilon_2^*} \sin\varepsilon^*_1 \right)
,
\end{multline}
which is minimum phase.
Similarly, for $G_{V_2,P_m}(s)$
\begin{multline}
\label{eq:zeroPmvolt}
\text{\normalsize$q^{(V_2,P_m)}$}
\! = 
\frac{-1}{T_{do}' b_\Delta}\left(b_\Delta +
b_\Sigma \left[1+\frac{E_N}{E_q'^*} \frac{\cos \varepsilon_2^*}{\sin \varepsilon_2^*}
\sin\delta^*\right]
\right)
\\
-\frac{K_A}{T_{do}'}\beta_1 \left(\cos\varepsilon^*_1 - \frac{\cos \varepsilon_2^*}{\sin \varepsilon_2^*} \sin\varepsilon^*_1 \right).
\end{multline}
With high load angles, and if the voltage $V_2$ is measured closer to the machine node than the machine terminal $|\varepsilon_2^*| < |\varepsilon_1^*| $, $q^{(V_2,P_m)}$ may potentially move into the right half-plane. 

\subsection{Active Power Injection and Phase Angle Measurement}
\label{sec:Pin_ctrl}
Power electronic devises can be used to improve the stability of electromechanical modes. If fast and strong enough, power oscillations can be controlled directly by controlling active power injections. The difference from the governor control in \cref{sec:aux_gov_ctrl} is that the input is not directly acting on the state $\omega$.
The transfer function to system states 
\begin{equation}
\label{eq:P_to_states}
(sI-A)^\inv \begin{bmatrix}
0\\ b_{2}\\ b_{3}
\end{bmatrix} = \frac{1}{p(s)} \begin{bmatrix}
(s+a_{33})b_{2}-a_{23}b_{3}\\s (s+a_{33})b_{2}-sa_{23}b_{3}\\ -  a_{31} b_{2} + (s^2 + a_{21})b_{3}
\end{bmatrix},
\end{equation}
is  a combination of \cref{eq:Ef_to_states,eq:Pm_to_states} 
where from \eqref{app:inputmatrix}
\begingroup\makeatletter\def\f@size{9}\check@mathfonts
\def\maketag@@@#1{\hbox{\m@th\normalsize\normalfont#1}}%
\begin{equation}
\label{eq:Bvariables}
b_{2} =  \frac{ \beta_3 E_{\subq}'^*}{MV_3^*}\cos\varepsilon^*_3, \ \text{\normalsize and} \ \
b_{3} =  \frac{\beta_3}{T_{do}' b_\Delta V_3^*} \left(\sin\varepsilon^*_3 -b^\text{AVR}_3\right) \!.
\end{equation}
\endgroup
The term $b^\text{AVR}_3$ in \cref{eq:Bvariables} is introduced by the AVR due to direct feed-through between the input bus and the machine terminal as shown in \eqref{eq:app_B_AVR_addition}.
\begin{rem_}
	The input matrix elements, $b_2$ and $b_3$ in \eqref{eq:Bvariables}, are not to be confused with susceptance. 
\end{rem_}
From calculations in \cref{app:direct_feedthorugh} we find that
\begin{equation}
\label{eq:bAVR}
b^\text{AVR}_3 =  \frac{b_\Delta K_A}{\beta_3 \bslash_{13}}\sin \varepsilon_{13}^*, 
\end{equation}
where $\bslash_{13} = b_{11}'+\smash{\frac{b_{11}'b_{3N}}{b_{13}}} + b_{3N}$ and $\varepsilon_{13}^* = {\theta_1^*-\theta_3^*}$. Thus 
\begin{equation}
\label{eq:b3_withAVR}
b_3 = 
\frac{\beta_3}{T_{do}' b_\Delta V_3^*}
\left(\sin \varepsilon_3^* - \frac{b_\Delta K_A}{\beta_3 \bslash_{13}}\sin \varepsilon_{13}^*\right)
.
\end{equation}

The transfer function $G_{\theta_2,P_3}(s)$ becomes $G_{\theta_2,P_3}(s)=$
\begingroup\makeatletter\def\f@size{9}\check@mathfonts
\def\maketag@@@#1{\hbox{\m@th\normalsize\normalfont#1}}%
\begin{equation}
\label{eq:transfer_funcP3toth2}
\frac{c_1[ (s\!+\!a_{33})b_2 \!-\!a_{23}b_3]\!+\! c_3 [(s^2 \!\!+ \!a_{21})b_{3} \!-\!  a_{31}b_{2}]+  d p(s)}{p(s)},
\end{equation}
\endgroup
where from \eqref{eq:SMIBout1}
\begin{equation}
\label{eq:Cvariables}
c_{1} =  \beta_2\frac{E_{\subq}'^*}{V^*_2}\cos\varepsilon^*_2, \quad 
c_{3} =  \beta_2\frac{1}{V^*_2}\sin\varepsilon^*_2,
\end{equation}
and from \cref{app:direct_feedthorugh}, the direct feed-through term
\begin{equation}
\label{eq:Dvariables}
d = \frac{\cos\varepsilon_{23}^*}{\bslash_{23} V_2^*V_3^*}.
\end{equation}
Substituting \eqref{eq:characteristic_eq} in \eqref{eq:transfer_funcP3toth2}, the zero polynomial of $G_{\theta_2,P_3}(s)$ becomes
\begingroup\makeatletter\def\f@size{10}\check@mathfonts
\def\maketag@@@#1{\hbox{\m@th\normalsize\normalfont#1}}%
\begin{multline}
\label{eq:zero_poly_P_to_theta}
s^3 d{}  + s^2 ( d{}a_{33}+c_3b_3) + s (d{} a_{21} +c_1b_2) 
+ d{}a_{21} a_{33}  -  d{}a_{23} a_{31} \\ +c_1a_{33} b_2 - c_1a_{23}b_3 + c_3a_{21}b_3 - c_3a_{31}b_2,
\end{multline}
\endgroup
which similar to \eqref{eq:characteristic_eq}, for reasonable load angles and large $K_A$, is stable if
\begin{equation}
\label{eq:cond1}
(c_3b_2+d{}a_{23})(c_1b_3 + d{}a_{31}) > 0.
\end{equation}
Under \Cref{asm:uniform_power_flow}, all load angles have the same sign, so \eqref{eq:cond1} reduces to
\begin{equation}
\label{eq:cond2}
\sin \delta^* (c_1b_3 + d{}a_{31}) > 0.
\end{equation}
Combining \eqref{eq:b3_withAVR} with \eqref{eq:Cvariables}, 
\begingroup\makeatletter\def\f@size{9.5}\check@mathfonts
\def\maketag@@@#1{\hbox{\m@th\normalsize\normalfont#1}}%
\begin{equation}
\label{c1b3}
c_1b_3 = \beta_2\frac{E_{\subq}'^*}{V^*_2}\cos\varepsilon^*_2 
\frac{1}{T_{do}' V_3^*}\left(\frac{\beta_3}{b_\Delta} \sin \varepsilon_3^* - \frac{K_A}{\bslash_{13}}\sin \varepsilon_{13}^*\right) 
\end{equation}
\endgroup
and \eqref{eq:Avariables} with \eqref{eq:Dvariables}, 
\begin{equation}
d{}a_{31} = \frac{\cos\varepsilon_{23}^*}{\bslash_{23} V_2^*V_3^*}
\left(\frac{b_\Sigma}{T_{do}' b_\Delta} E_N \sin\delta^* - \frac{K_A}{T_{do}'}\beta_1 E_{\subq}'^\steady\sin\varepsilon^*_1\right).
\end{equation}

It now follows that if \eqref{eq:destabCond} is fulfilled, then $G_{\theta_2,P_3}(s)$ is minimum phase. 
Conditions \cref{eq:destabCond,eq:cond2} are similar but with an extra term from 
\eqref{c1b3} that relaxes the condition in \eqref{eq:cond2} as long as $|\smash{{\beta_3} \sin \varepsilon_3^*/{b_\Delta} - {K_A}\sin \varepsilon_{13}^*/{\bslash_{13}}}|>0$, which holds true for $\varepsilon_{13}^*\sign(\delta^*)<0$ and for small $|\varepsilon_{13}^*|\ll|\varepsilon_{3}^*|$. The system can therefore be minimum phase even if it is unstable.

\begin{rem_}
	The zero of interest in $G_{\theta_2,P_3}(s)$ is a complex conjugated zero pair associated with the electromechanical dynamics of the rotor. 
	The undamped frequency of the electromechanical mode \eqref{eq:OMEGA} is $\Omega = \smash{\sqrt{{b_\Sigma}E_{\subq}'^*E_N\cos\delta^*\text{/}M}}$. Assuming that interaction between the electromechanical and voltage dynamics can be neglected, i.e. if we have low load angles and no AVR, then the
	electromechanical zero pair is given by, $\pm \smash{j\sqrt{a_{21} + c_1b_2\text{/}d}}$ 
	assuming that $E_\subq'\approx E_N$, $\varepsilon_{2}^* \approx \varepsilon_{3}^*$, and  $\cos \delta^* \approx \cos \varepsilon_2^*\cos \varepsilon_3^*$ then
	\begin{equation}
	\Imag q_{1,2}^{(\theta_2,P_3)} \approx \pm \sqrt{\frac{b_{12}'}{M}E_q'^{*2} \cos \delta^* }, \quad \text{where} \quad \left|q_{1,2}^{(\theta_2,P_3)}\right|\geq \Omega.
	\end{equation}
	The closer the control and measurement are to the machine node, the faster the zero. The mode is unobservable at the infinite bus, where $b_{12}' = b_\Sigma$.
\end{rem_}

\begin{figure}[t]
	\centering
	\includegraphics[scale=0.9]{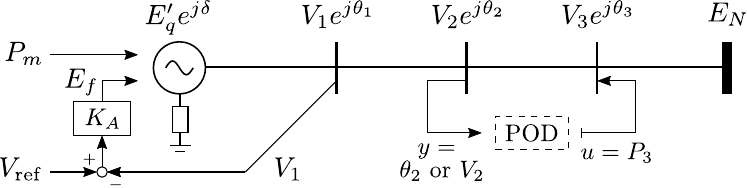} 
	\caption{SMIB model with POD controller.}
	\label{fig:SMIBpod}
\end{figure}
\subsection{Active Power Injection and Voltage Measurement}
\label{sec:Pin_ctrl_voltage}
Consider active power injections as in \Cref{sec:Pin_ctrl}, but with voltage amplitude at a network node as the measured output. The zero polynomial in the transfer function from  $P_3$ to $V_2$ will be the same as \eqref{eq:zero_poly_P_to_theta} but with
\begin{equation}
\label{eq:Cvariables_VOLT}
c'_{1} =  -\beta_2{E_{\subq}'^*}\sin\varepsilon^*_2, \quad 
c'_{3} =  \beta_2\cos\varepsilon^*_2,
\end{equation}
and  direct feed-through term
\begin{equation}
\label{eq:direct_P_to_V}
d'{} = \frac{\sin\varepsilon_{23}^*}{\bslash_{23} V_3^*}.
\end{equation}
Here the step from \eqref{eq:cond1} to \eqref{eq:cond2} is no longer valid. 
To analyze the presence of NMP zeros, we first make the following assumptions.
\begin{asm}[$\theta_2=\theta_3$]
	\label{asm:coinciding_nodes1}
	Control and measurement occur at the same bus. Therefore, the direct term \eqref{eq:direct_P_to_V} is zero. 
\end{asm}
The zero polynomial of $G_{V_2,P_3}(s)$ become 
\begin{equation}
\label{eq:zero_poly_P_to_V1}
s^2{c'_3b_3}  + s {c'_1b_2 }+ {c'_1(a_{33} b_2 - a_{23}b_3 )+ c'_3(a_{21}b_3 - a_{31}b_2)},
\end{equation}
which divided by $c'_3 b_3$ gives us the zero polynomial
\begin{equation}
\label{eq:zero_poly_P_to_V}
s^2  + s \alpha_1 + \alpha_2.
\end{equation}
\begin{asm}[$\theta_1=\theta_2=\theta_3$]
	\label{asm:coinciding_nodes2}
	Control and measurement both occur at the machine terminal.
\end{asm}
Substituting \cref{eq:Bvariables,eq:b3_withAVR,eq:Cvariables_VOLT} in \eqref{eq:zero_poly_P_to_V} we find that
\begin{equation}
\alpha_1
=- \frac{E_q'^{*2} T_{do}' b_\Delta \sin \varepsilon_2^*\cos \varepsilon_3^*}{M \cos \varepsilon_2^* \left(\sin \varepsilon_3^* - \frac{b_\Delta K_A}{\beta_3 \bslash_{13}}\sin \varepsilon_{13}^*\right)} 
= - \frac{E_q'^{*2} T_{do}'b_\Delta}{M}
\end{equation}
where the 2\textsuperscript{nd} equality follows due to \Cref{asm:coinciding_nodes2}. As shown in \Cref{app:coef_alpha2}, \Cref{asm:coinciding_nodes2} also mean that
\begin{equation}
\alpha_2 \approx -\frac{E_q'^{*2} (b_\Sigma+b_\Delta)}{M}.
\end{equation}
\begin{asm}[$|\alpha_1|\gg |\alpha_2|$]
	\label{asm:Tdo=10}
	With parameters in \si{\perunit} it is reasonable that $T'_{do} \approx \SI{10}{\second}$ and that $b_\Delta > b_\Sigma$ \citep{KundurPowerSystemStability1994}. Therefore, $\alpha_1$ dominates $\alpha_2$ in  \cref{eq:zero_poly_P_to_V}.
\end{asm}
Under  \Cref{asm:Tdo=10}, the transfer function $G_{V_1,P_1}(s)$ have an open-loop NMP zero at $-\alpha_1$.
Note  also that the undamped frequency of the electromechanical mode 
\begin{equation}
\label{eq:greaterThanOmega}
\Omega = \sqrt{\frac{b_\Sigma}{M} E_{\subq}'^\steady E_N \cos\delta^*} \ll \frac{E_q'^{*2} T_{do}'b_\Delta}{M}.
\end{equation}
Thus, the NMP zero does not indicate any damping control limitations.
If we relax \Cref{asm:coinciding_nodes2} but still assume that $\alpha_1$ dominates $\alpha_2$, then $G_{V_i,P_i}(s)$, $i=1,2,3$, has an open-loop zero
\begin{equation}
\label{eq:q_Vi_Pi}
q^{(V_i, P_i )}\approx \frac{E_q'^{*2} T_{do}' b_\Delta \sin \varepsilon_i^*}{M  \left(\sin \varepsilon_i^* - \frac{b_\Delta K_A}{\beta_i\bslash_{1i}}\sin \varepsilon_{1i}^*\right)}.
\end{equation}
\begin{rem_}
	If $\varepsilon_{1i}^*\sign(\delta^*) <0$ the NMP zero moves towards the origin. This may limit damping control design. If  $\varepsilon_{1i}^*\sign(\delta^*)  >0$ the zero moves further into the right half-plane and eventually crosses over into the left half-plane where it goes towards the origin. 
	Typically however, $|\varepsilon_i^*| > |\varepsilon_{1i}^*| \approx 0$ and thus \eqref{eq:q_Vi_Pi} is insensitive to both AVR and the control bus location.
\end{rem_}
If we relax \Cref{asm:coinciding_nodes1} and allow $\theta_2\neq \theta_3$ then $G_{V_2,P_3}(s)$ has an open-loop right half-plane zero
\begin{equation}
\label{eq:q_V2_P3}
q^{(V_2, P_3 )}\approx \frac{E_q'^{*2} T_{do}' b_\Delta }{M}\frac{\tan \varepsilon_2^*}{\tan \varepsilon_3^*}.
\end{equation}
Thus if the measurement bus is closer to the machine than the control bus, the NMP zero moves closer to the origin. 

\subsection{Summary}
In this section, fundamental control limitations in a SMIB power system have been analyzed by studying the presence of NMP zeros in open-loop transfer functions. 

The $P$-$\theta$ transfer function $G_{\theta_2,P_3}(s)$  has a zero pair 
\begin{equation}
q_{1,2}^{(\theta_2,P_3)} \approx \sigma \pm j \sqrt{\frac{b_{12}'}{M}E_q'^{*2} \cos \delta^* }.
\end{equation}
The condition for $\sigma <0$ is that
\begin{equation}
\label{eq:destabCondSum}
\left|\frac{T_{do}'}{d}c_1b_3 + \frac{b_\Sigma}{b_\Delta} E_N \sin\delta^*\right| >
\left|K_A\beta_1 E_{\subq}'^\steady\sin\varepsilon_1^*\right|,
\end{equation}
where $c_1b_3$ from \eqref{c1b3} depends on the location of the measurement and control bus respectively.
The sign of $\sigma$ is closely linked to the destabilizing effect that the AVR has on the electromechanical mode. The closer the control and measurement is to the machine node, however, the less sensitive the system is to the effect of the AVR.

The $P$-$V$ transfer function is less sensitive to the AVR. Its NMP zero \eqref{eq:q_V2_P3}
tells us that control input should preferably be close to the machine node and that the measurement is best located further out in the system.  This agrees with \cite{smedUtilizingHVDCDamp1993} where active power controllability and phase angle observability was found most effective far away from mass-weighted electrical midpoint, which for the SMIB model means far away from the infinite bus.
On the contrary, reactive power controllability and voltage observability is shown to be the most effective at the midpoint. However, since the voltage at the infinity bus is assumed fixed, this makes the SMIB model unsuitable for the study of $Q$-$V$ control, as noted in \Cref{rem:Vref_to_V2}. 

\section{Simulation Study}
\label{sec:simulation}
In this section we study the control limitations imposed by zero dynamics using a more detailed power system model implemented in Simulink. The considered SMIB system shown in \cref{fig:kundursmib} has a 6\textsuperscript{th} order synchronous machine model and fast, but not neglected, exciter dynamics. The model is described in detail in \cite[Example~13.2]{KundurPowerSystemStability1994}, where it is used to study the effect of AVR and PSS. \color{black}
The machine, representing the aggregation of four synchronous machines, feeds \SI{0.9}{\perunit} active power into an infinite bus. 
\subsection{Active Power Injection and Phase Angle Measurement}
\label{sec:simul_theta}
\begin{figure}[t]
	\centering
	\includegraphics[scale=0.9]{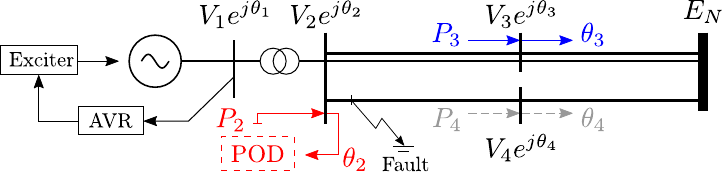}
	\caption{SMIB network from \cite[Example~13.2]{KundurPowerSystemStability1994}.}
	\label{fig:kundursmib}
\end{figure}

To test the transient and steady-state rotor angle stability, we consider a three phase ground fault at time $t=\SI{1}{s}$. The fault occurs in the lower circuit close to bus 2 as shown in \cref{fig:kundursmib}.
The fault is cleared  by disconnecting the affected line at both ends within \SI{0.10}{\second}. 

\begin{figure}[t]
	\centering
	\includegraphics[width=\linewidth]{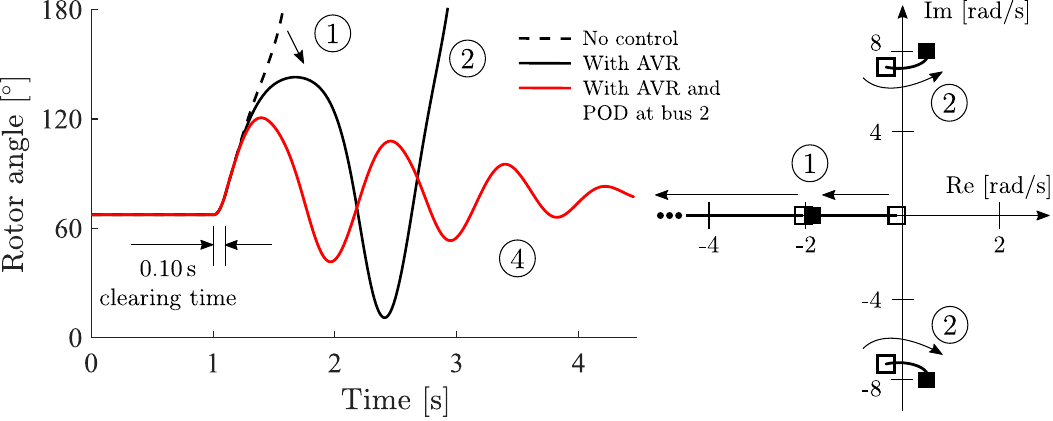}
	\begin{minipage}{0.53\linewidth}
		\centering
		\caption{Rotor angle response with fault cleared in \SI{0.10}{\second}.}
		\label{fig:Kundur_a}
	\end{minipage}
	\hfill
	\begin{minipage}{0.39\linewidth}
		\caption{Root locus with AVR at bus 1.}
		\label{fig:Kundur_b}		
	\end{minipage}
\end{figure}

\begin{figure}[t]
	\centering
	\begin{subfigure}[b]{0.5\linewidth}
		\includegraphics[scale=0.98]{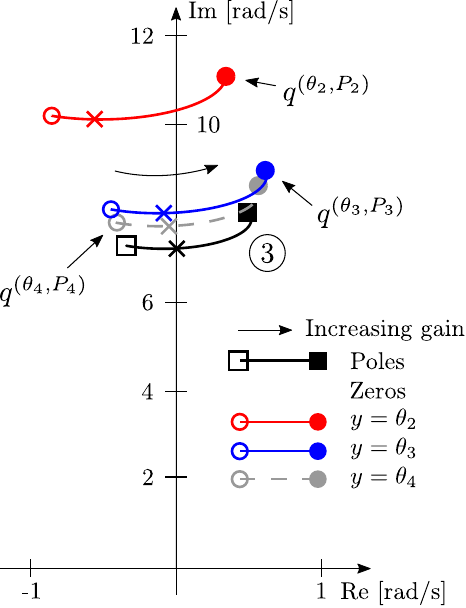}
		\caption{AVR at bus 1.}
		\label{fig:Kundur_c}
	\end{subfigure}%
	\begin{subfigure}[b]{0.5\linewidth}
		\hfill
		\includegraphics[scale=0.94]{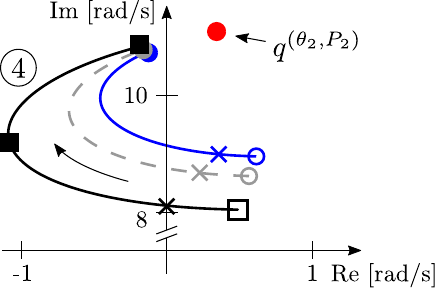}
		\caption{POD at bus 2.}
		\label{fig:Kundur_d}
		\vspace{3pt}
		\hfill 
		\includegraphics[scale=0.94]{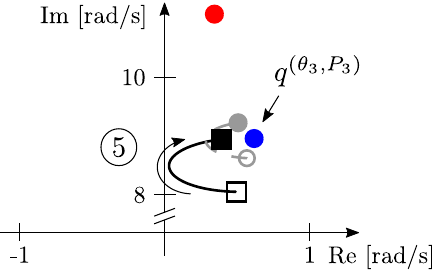}
		\caption{POD at bus 3.}
		\label{fig:Kundur_e}
	\end{subfigure}%
	\caption{
		Poles and zeros of $ G_{\theta_i,P_i}(s)$, $i =2,3,4 $, with AVR or AVR plus POD control. The ``$\times$'' marks the feedback gain level where the pole crosses the imaginary axis.}
	\label{fig:rlocus2}
\end{figure}

Following the numbers listed in \cref{fig:Kundur_a,fig:Kundur_b,fig:rlocus2}.
\begin{enumerate}	
	\Myitem With constant field voltage, the system fails to maintain synchrony following the fault in \cref{fig:Kundur_a}. To enhance the transient stability, AVR is implemented. As seen in \cref{fig:Kundur_b} this moves the poles associated to the voltage dynamics further into the LHP, increasing the synchronizing torque of the machine. 
\end{enumerate}	
The presence of an extra pole on the real axis in \cref{fig:Kundur_b} stems from the fact that the electrical dynamics of the synchronous machine is represented by a 6\textsuperscript{th} order model.
\begin{enumerate}
	\setcounter{enumi}{1}
	\Myitem Although the AVR achieves initial transient stability the system goes unstable in the second swing in \cref{fig:Kundur_a}. This is because the AVR has moved the pole of the electromechanical mode into the right half-plane as seen in \cref{fig:Kundur_b}. 
\end{enumerate}
To stabilize the system we consider active power injections using local phase angle measurements at buses 2--3 in \cref{fig:kundursmib}. 
\begin{enumerate}
	\setcounter{enumi}{2}
	\Myitem As shown in \cref{sec:Pin_ctrl}, the presence of NMP zeros are closely linked to the destabilizing effect of the AVR. For nodes closer to the machine, the zeros are shifted further into the left half-plane. This can be seen in \cref{fig:Kundur_c} where the open-loop zeros are plotted alongside the electromechanical mode.
\end{enumerate}
We consider a classical POD design using the residue method. Let $
P_i = -K(s) \theta_i
$,
where the feedback controller
\begin{equation}
\label{eq:controller_simulink}
K(s) = 
\underbrace{\frac{s+T_1}{s+\smash{T_2}}}_{\mathclap{\text{Phase compensation}\hspace{6mm}}}  s \,
\bigg(\underbrace{\frac{100}{s+100}}_{\text{Low-pass}} 
\bigg)^{\mathclap{2}}
\underbrace{\frac{s}{s+\smash{1/1.5}}}_{\text{Washout}}
{\smash{K_\text{POD}}}.
\end{equation} 
The eigenvalue sensitivity to changes in $K(s)$ is given by  the residue \citep{pagolaSensitivitiesResiduesParticipations1989}
\begin{equation}
\label{eq:residue}
R(\lambda)= -\frac{\partial \lambda}{\partial K(s)}.
\end{equation}
The phase compensation in \eqref{eq:controller_simulink} is tuned so that $\arg R(\lambda) K(\lambda) = - \pi$ for the electromechanical mode. Thus, feedback moves the eigenvalue in the negative real direction as seen in \cref{fig:Kundur_d,fig:Kundur_e}. 
However, as the gain $K_\text{POD}$ increases, the trajectory of the closed-loop eigenvalue changes and it eventually approaches the position of the nearby open-loop zero.
\begin{enumerate}
	\setcounter{enumi}{3}
	\Myitem 
	With POD at bus 2, the control achieves an optimal damping of \SI{13}{\percent} as shown in \cref{fig:Kundur_d}. With POD and AVR, the system achieves both transient and steady-state stability as seen in \cref{fig:Kundur_a}. 
\end{enumerate}
Note that the  zero $\smash{q^{(\theta_2,P_2)}}$ does not move in \cref{fig:Kundur_d}. This is natural since closed-loop zeros in a SISO system cannot be moved by feedback control~\citep{seronFundamentalLimitationsFiltering1997}. 
\begin{enumerate}
	\setcounter{enumi}{4}
	\Myitem With an increasing feedback gain, the closed-loop poles approaches the open-loop zeros. In \cref{fig:Kundur_e} we see that this impose a limitation for control at bus~3 that fails to achieve stability for any $\smash{K_\text{POD}}$.
\end{enumerate}
The residue \eqref{eq:residue} is useful to find suitable input-output pairs as it is a measure of the controllability and observability of the considered mode \citep{pagolaSensitivitiesResiduesParticipations1989}. In general $\lambda$ can be stabilized as long as $|R(\lambda)|\neq 0$. With nearby NMP zeros however, robustness deteriorates, increasing the sensitivity to changes in the system \citep{seronFundamentalLimitationsFiltering1997}.
\subsection{Active Power Injection and Voltage Measurement}
\begin{figure}[t!]
	\centering
	\begin{subfigure}[b]{1\linewidth}
		\centering
		\includegraphics[scale=0.95]{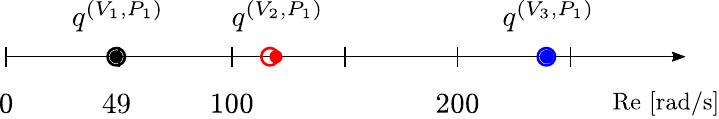}
		\caption{Active power injection at bus 1.}
		\label{fig:Kundur_Za}
		\vspace{6pt}
		\includegraphics[scale=0.95]{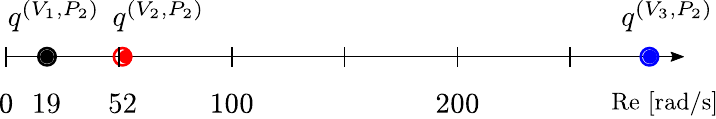}
		\caption{Active power injection at bus 2.}
		\label{fig:Kundur_Zb}
		\vspace{6pt}
		\includegraphics[scale=0.95]{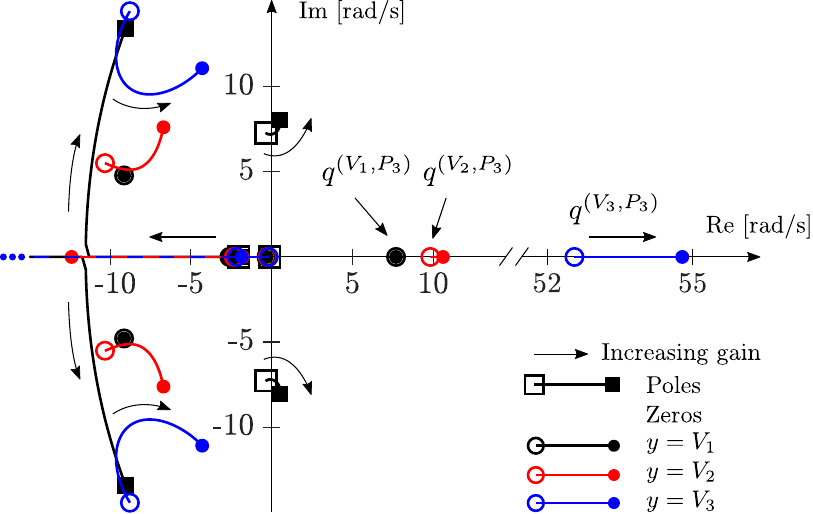}
		\caption{Active power injection at bus 3.}
		\label{fig:Kundur_Zc}
	\end{subfigure}%
	\caption{
		Poles and zeros of $G_{V_i,P_k}(s)$, $i=1,2,3$, with AVR.}
	\label{fig:rlocus3}
\end{figure}

Consider the system in \cref{fig:kundursmib} as in \cref{sec:simul_theta} but now with voltage as the measured output. According to \eqref{eq:q_V2_P3}, the transfer function $G_{V_i,P_k}(s)$ should have a zero on the positive real axis. The zero should be roughly at the same point for all local measurement loops $G_{V_i,P_i}(s)$. In \cref{fig:rlocus3} we see that these zeros appear at \SI{50}{\radian\per\second} which, in agreement with \eqref{eq:greaterThanOmega}, is much larger than the electromechanical mode. With external measurement, the NMP zero moves closer to the origin if the measurement is closer to the machine and vice versa. As shown in \cref{sec:Pin_ctrl_voltage}, all NMP zeros are insensitive to the AVR. 
\section{Conclusions}
\label{sec:Conclusions}
The presence of open-loop transfer function zeros have been characterized for different input-output configurations in power systems. It was shown, using a SMIB model, that NMP zeros are closely linked to the destabilizing effect of AVR. Depending on input-output combination chosen for feedback control, these NMP zeros may persist in the system. Right half-plane zeros close to an unstable electromechanical mode was shown to prevent stabilization using feedback control.

The model detail will of course have an impact on the pole-zero locations of the system. For instance, the approximation from the dynamical AVR model \eqref{eq:AVR_original} to the simple proportional model \eqref{eq:AVR} is only accurate if the AVR is fast compared to the electromechanical dynamics. However, reducing the bandwidth of the AVR also means that the intended transient stability improvement will be reduced. So there will still be a trade-off between transient and small-signal stability as described in \cref{sec:AVRstab}.

The SMIB model have been used since it allows for an analytically tractable solution to the problem. Future work will extend this analysis to general multi-machine systems. Here there will be more dynamical interactions and multiple generators participating to various degrees in the electromechanical modes. Consequently, zeros associated with the poorly damped electromechanical modes are likely to be present in most input-output combinations.

\bibliography{refs}                                               
\appendix
\section{Direct Feed-Through}
\label{app:direct_feedthorugh}
The admittance matrix interconnecting network nodes $i$ and $k$  in \cref{fig:SMIB} is given by 
\begingroup\makeatletter\def\f@size{10}\check@mathfonts
\def\maketag@@@#1{\hbox{\m@th\normalsize\normalfont#1}}%
\begin{equation}
\setlength\arraycolsep{1pt}
Y_{\theta\theta} = j\begin{bmatrix}
-b_{1i}'-b_{ik} & b_{ik}
\\
 b_{ik} & -b_{ik}-b_{kN} 
\end{bmatrix},
\end{equation}
\endgroup
with nodes ordered so that $b_{1i}'\geq b_{1k}'$, i.e., node $i$ is closer to the machine node.
By \eqref{def:weightAdmit}, the weighted admittance matrix 
\begin{equation}
\Ybus_{\theta\theta} =
j\begin{bmatrix}
(b_{1i}'+b_{ik})V_i^2 & -b_{ik}V_iV_ke^{j\varepsilon_{ik}}
\\
-b_{ik}V_iV_ke^{-j\varepsilon_{ik}} & (b_{ik}+b_{kN})V_k^2
\end{bmatrix}, 
\end{equation}
where $\varepsilon_{ik} = \theta_i-\theta_k$. With $\Ybus_D = $
\begin{equation}
\smash{-\Ybus_{\theta\theta}^\inv} =
j\frac{1}{V_i^2V_k^2}\frac{1/b_{ik}}{\bslash_{ik}}\begin{bmatrix}
(b_{ik}+b_{kN})V_k^2 & b_{ik}V_iV_ke^{j\varepsilon_{ik}}\\
b_{ik}V_iV_ke^{-j\varepsilon_{ik}} & (b_{1i}'+b_{ik})V_i^2
\end{bmatrix},
\end{equation}
where $\bslash_{ik} = b_{1i}'+{\frac{b_{1i}'b_{kN}}{\smash{b_{ik}}}} + b_{kN}$, the direct feed-through between the network nodes are then obtained using \eqref{app:feedthroughmatrix}.

\section{Zero Polynomial Coefficient}
\label{app:coef_alpha2}
Substituting \cref{eq:Avariables,eq:Bvariables,eq:b3_withAVR,eq:Cvariables_VOLT} in \eqref{eq:zero_poly_P_to_V} we find that
\begingroup\makeatletter\def\f@size{9}\check@mathfonts
\def\maketag@@@#1{\hbox{\m@th\normalsize\normalfont#1}}%
\begin{multline}
\text{\normalsize
$\alpha_2 = \big[{ - c_1'a_{23}b_3 + c_3'a_{21}b_3 + c_1'a_{33} b_2- c_3'a_{31}b_2}\big]/c_3'b_3 
=$}
\\
 \frac{E_q'^{*2} b_\Sigma}{M } \frac{1}{\cos \varepsilon_2^* \left(\sin \varepsilon_3^* -
 	b_3^\text{AVR}
 	\right)} 
\bigg[
\frac{E_N}{E_q'^*} 
\sin \delta^* \sin \varepsilon_2^*  \left(\sin \varepsilon_3^* -
b_3^\text{AVR}
\right) 
\\
+{\frac{{E_N}}{E_\subq'^*}}
\cos\delta^* \cos\varepsilon_2^* \left(\sin \varepsilon_3^* -
b_3^\text{AVR}
\right)-{\frac{b_\Sigma+b_\Delta}{b_\Sigma}} \sin \varepsilon_2^* \cos \varepsilon_3^*
\\
-{\frac{b_\Delta}{{b_\Sigma}}}K_A\beta_1 \cos \varepsilon_1^* \sin \varepsilon_2^* \cos \varepsilon_3^*
-{\frac{{E_N}}{E_\subq'^*}}\sin\delta^*\cos\varepsilon_2^*\cos\varepsilon_3^*
\\
 +\frac{b_\Delta}{{b_\Sigma}}K_A\beta_1 \sin \varepsilon_1^* \cos \varepsilon_2^* \cos \varepsilon_3^*
 \smash{\bigg]} 
{\quad\approx-\frac{E_q'^{*2} (b_\Sigma+b_\Delta)}{M} }
,
\end{multline}
\endgroup
where from \eqref{eq:bAVR}, $b_3^\text{AVR} = {\frac{b_\Delta K_A}{{\beta_3 \bslash_{13}}} \vspace{2em}}  \sin \varepsilon_{13}^*$.
Note that all effects from the AVR cancel out due to \Cref{asm:coinciding_nodes2}.

\end{document}